\documentclass[nofootinbib,notitlepage,superscriptaddress,10pt,aps,pra]{revtex4-1}
\usepackage{amssymb}

\usepackage{amsmath}

\usepackage{graphicx}

\usepackage{amsfonts}

\begin{document}

\title{%
Relative Locality in Curved Space-time}
\author{Jerzy Kowalski-Glikman}\email{Jerzy.Kowalski-Glikman@ift.uni.wroc.pl}\affiliation{Institute
for Theoretical Physics, University of Wroc\l{}aw, Pl.\ Maksa Borna
9, Pl--50-204 Wroc\l{}aw, Poland}
\author{Giacomo Rosati}\email{Giacomo.Rosati@ift.uni.wroc.pl}\affiliation{Institute
for Theoretical Physics, University of Wroc\l{}aw, Pl.\ Maksa Borna
9, Pl--50-204 Wroc\l{}aw, Poland}

\begin{abstract}
In this paper we construct the action describing dynamics of the
particle moving in curved spacetime, with a non-trivial momentum
space geometry. 
Curved momentum space is the core feature of 
theories where relative locality effects are presents.
So far aspects of nonlinearities in momentum space 
have been studied only for flat or constantly expanding (De Sitter) spacetimes, 
relying on the their maximally symmetric nature.
The extension of curved momentum space frameworks to arbitrary 
spacetime geometries could be relevant for the opportunities to test
Planck-scale curvature/deformation of particles momentum space.
As a first example of this construction we describe the
particle with kappa-Poincar\'e momentum space on a circular orbit in
Schwarzschild spacetime, where the contributes of momentum space curvature turn out to be negligible. The analysis of this problem relies crucially on the solution of the soccer ball problem.
\end{abstract}

\maketitle

It is widely expected that quantum gravity effects are characterized
by the presence of an energy scale, usually identified with the
Planck mass $M_{Pl} = \sqrt{G/\hbar}\sim 10^{19}$ GeV, and the
associated length scale, the Planck length $\ell_{Pl}
=\sqrt{G\,\hbar}\sim 10^{-35}$ m. Unfortunately, processes with
characteristic energy and length  of order of the Planck scales are
far beyond reach of the current, and foreseeable, technologies and
therefore the question arises as to if there could be some traces of
quantum gravity effects observable with the current, or near future
experiments.

As advocated in \cite{AmelinoCamelia:2011bm} such an opportunity
might be offered by a possible semiclassical, weak gravity regime of
quantum gravity, in which both $G$ and $\hbar$ are very small, while
their ratio remains finite. In such a regime the dynamics of
particles and fields depends on a mass scale $\kappa$, which in the
case of elementary systems could be identified with the Planck mass.
Further it is argued that the presence of the mass scale has an
effect similar to the one well known in 2+1 gravity~\cite{deSousaGerbert:1990yp,Matschull:1997du,Meusburger:2005mg,Schroers:2007ey}, namely that the momentum space becomes `curved'\footnote{The idea of a curved momentum space appeared already in Refs.~\cite{born} and~\cite{Snyder:1946qz}, and was later developed by the Russian group (see Ref.~\cite{Kadyshevsky:1977mu} and references therein).}, i.e., starts having a non-trivial geometry, with the characteristic scale $\kappa$~\cite{Majid:1999tc,Magueijo:2001cr,KowalskiGlikman:2002ft,KowalskiGlikman:2003we,Oriti:2009wn,Arzano:2010jw,Chang:2010ir,AmelinoCamelia:2011bm}. See \cite{Kowalski-Glikman:2013rxa} for the up to date review.

Since the non-trivial geometry of momentum space almost inevitably
leads to the conflict with absolute spacetime locality, the class of
theories with curved momentum space are characterized also by the
necessary controllable relaxation of the absolute locality~\cite{bob}, \cite{k-bob},
resulting in the so-called relative locality principle
\cite{AmelinoCamelia:2011bm}, \cite{AmelinoCamelia:2011pe}.

Although the relative locality principle, as originally devised, arises in the
limit $G\rightarrow0$ and therefore seems to force us to restrict
our attention to the flat Minkowski space with vanishing gravity, it
is possible to argue that one can still consider the models of
particles and fields having curved momentum space and evolving in a
nontrivially curved spacetime. First, even if in the limiting
procedure leading to the curved momentum space the Newton's constant
is assumed to be small, the physical effect of classical gravity,
which are of order of $GM$, where $M$ is the mass/energy of the
source, might still be not negligible. Secondly, as in 2+1 gravity,
it could be that the curvature of momentum space results from the
`topological' sector of gravity (see \cite{Kowalski-Glikman:2013rxa}
for discussion and references) which is superimposed on the
dynamical gravity sector which exhibits itself in the spacetime
curvature.

The theory of motion of particles in de Sitter spacetime with non
linearities in momentum space has been discussed in
\cite{DSR-DeSitter}. In the analysis reported there, the constant
expansion rate is introduced, in a theory with deformed relativistic
symmetries \cite{AmelinoCamelia:2000ge},
\cite{AmelinoCamelia:2000mn}, as a third observer-invariant scale
(besides $c$ and $\kappa$; see also \cite{KowalskiGlikman:2004kp}.)

In this paper we present the general construction of the action of a
relativistic particle moving in spacetime of an arbitrary geometry
encapsulated in the spacetime tetrad $e^a_\mu(x)$ with curved
momentum space characterized by the tetrad $E_a^\alpha(p)$.
One can show that our construction is compatible with 
the examples of relative locality studies present in the literature 
(e.g.~\cite{bob},~\cite{k-bob},~\cite{DSR-DeSitter}), as well as with 
the analyses based on the principle of relative locality (e.g.~\cite{AmelinoCamelia:2011bm},~\cite{anatomy},~\cite{spinning}).

Let us start our construction with the standard action of
relativistic particle moving in flat Minkowski space. It reads
\begin{equation}\label{1}
 S^{0} = \int\, d\tau\, \frac{d x^a}{d\tau}\,
    p_a - N(p^ap_a+m^2)\,,
\end{equation}
where $N$ is the Lagrange multiplier enforcing the mass shell
condition and we rise and lower indices in spacetime and momentum
space with Minkowski metric tensor $\eta_{ab}$. In fact $N$ is also
a gauge field associated with the local reparametrization invariance
on the worldline $\tau\rightarrow\tau'(\tau)$ and therefore it can be
gauge fixed to an appropriate (usually constant) value. The
equations of motions following from (\ref{1}) are
\begin{equation}\label{2}
    \dot p =0\,,\quad \dot x^a =2N\, p^a\,,\quad p^2=-m^2\,,
\end{equation}
and to get the standard velocity-momentum relation we gauge fix
$N=1/2m$.

The action (\ref{1}) can be generalized so as to describe the
particle moving in curved spacetime or having curved momentum space.
In the first case we introduce the spacetime tetrad $e_\mu^a(x)$
that maps the tangent space of the curved spacetime manifold to an
ambient Minkowski space. The action has the form
\begin{equation}\label{3}
 S^{E} =\int\, d\tau\,  \frac{d x^\mu}{d\tau}\,e_\mu^a(x)
  \, p_a - N(p^ap_a+m^2)\,.
\end{equation}
Since this action is a bit unusual let us stop here to check that
it leads to the geodesic equation as the standard action $\int
d\tau\sqrt{|g_{\mu\nu}\, \dot x^\mu \dot x^\nu|}$ does. The equations
of motion following from (\ref{3}) are (for $N$ gauge fixed to
$1/2\, m$, as before)
\begin{equation}\label{4}
    \dot x^\mu\, e^a_\mu = \frac1m\, p^a\,,\quad
    (e^a_{\nu,\mu}-e^a_{\mu,\nu})\,
    \dot x^\nu \, p_a -e^a_\mu\, \dot p_a =0\,,\quad p^2+m^2=0\,,
\end{equation}
Substituting the momentum from the first equation to the second,
after straightforward algebra one derives the geodesic equation.

Turning things around we can write down the action for a particle
with curved momentum space \cite{AmelinoCamelia:2011bm},
\cite{Kowalski-Glikman:2013rxa}
\begin{equation}\label{5}
 S^{\kappa} =-\int\, d\tau\,  x^a\,E^\alpha_a(p)
  \,\frac{d  p_\alpha}{d\tau} + N({\cal C}(p)+m^2)\,,
\end{equation}
where the function ${\cal C}(p)$ in the mass shell relation is
defined to be the geodesic distance form the momentum space origin
to the point with coordinates $p_\alpha$. The equations of motion
following from (\ref{5}) are
\begin{equation}\label{6}
    \dot x^a\, E_a^\alpha = N\, \frac{\partial {\cal C}(p)}{\partial p_\alpha}\,,\quad
     \dot p_\alpha =0\,,\quad {\cal C}(p)+m^2=0\,.
\end{equation}
where in the first equations we omitted the terms proportional to
$\dot p_\alpha$.

 Before making the decisive step, formulating the
action which combines  (\ref{3}) and (\ref{5}) let us notice that
for curved spacetime the positions $x^\mu$ have the upper index
(chosen from the middle of Greek alphabet $\mu$, $\nu$, \dots),
while for curved momentum space the momenta $p_\alpha$ have the
lower index (chosen from the beginning of Greek alphabet $\alpha$,
$\beta$, \dots). Therefore the spacetime tetrad mapping from the
tangent space of spacetime into the ambient Minkowski space looks
like $e_\mu^a(x)$, while the momentum space one, mapping from the
tangent space of the momentum space into this same ambient Minkowski
space has the form $E^\alpha_a(p)$.

It follows that the only action for the particle with curved both
spacetime and momentum space, which reduces to (\ref{3}) or
(\ref{5}) when spacetime or momentum space are flat i.e.,
$e_\mu^a(x)=\delta_\mu^a$ or $E^\alpha_a(p)=\delta^\alpha_a$,
respectively has the form
\begin{equation}\label{7}
 S^{E\kappa} =-\int\, d\tau\,  x^\mu\,E^\alpha_a(p)
  \,\frac{d  }{d\tau} \left(e^a_\mu(x)\, p_\alpha\right)+ N({\cal
  C}(p)+m^2)\,.
\end{equation}
The equations of motion following from (\ref{7}) have the form
\begin{align}
 E_{a}^{\alpha} \frac{d}{ds} \left(e_{\mu}^{a}p_{\alpha}\right) &=  p_{\alpha}
e_{\nu,\mu}^{a} \frac{d}{ds} \left(x^{\nu} E_{a}^{\alpha} \right) \label{8a}\,,\\
 e_{\mu}^{a} \frac{d}{ds} \left( x^{\mu} E_{a}^{\alpha} \right) &= x^{\mu}
E_{a}^{\beta,\alpha} \frac{d}{ds} \left( e_{\mu}^{a} p_{\beta}
\right) + N \frac{
\partial {\cal C}(p)}{\partial p_{\alpha}}\label{8b}\,, \\
{\cal   C}(p) & = - m^2\label{8c}\,.
\end{align}

To see these equations in action let us consider a simple example of
a body on a circular orbit in Schwarzschild geometry, with
$\kappa$-Poincar\'e momentum space geometry  \cite{majrue}. In this
case (see e.g., \cite{Kowalski-Glikman:2013rxa} for discussion and
references,  \cite{anatomy} for an explicit analysis) the
non-vanishing components of the momentum space tetrad are
 \begin{equation}\label{9}
    E^0_{(0)} = 1\,,\quad E^i_{(j)} = e^{p_0/\kappa}\, \delta_i^j\,,\quad
    i,j=1,2,3\,,
 \end{equation}
where we use parentheses $(\cdot)$ to denote `flat' indices, and the
modified on-shell relation has the form
\begin{equation}\label{10}
    \cosh\frac{p_0}\kappa -\frac{
    \mathbf{p}^2}{2\kappa^2}\,e^{p_0/\kappa}=\cosh\frac{m}\kappa\,.
\end{equation}
We use the standard Schwarzschild spherical spacetime coordinates
$(t,r,\theta,\phi)$ in which the non-vanishing components of the
tetrad are
\begin{equation}\label{11}
    e_0^{(0)}=f(r), \quad e_1^{(1)} =
    f(r)^{-1}\,,\quad e_2^{(2)} =r\,,\quad e_3^{(3)}
    =r\sin\theta\,.
\end{equation}
with $f(r) = \sqrt{1-{2GM}/r}$.

We will be interested only in the circular, planar orbit, which
means that we will solve eqs.\ (\ref{8a})--(\ref{11}) supplemented
by the condition $\dot \theta=\dot r =0$. Taking these into account
from (\ref{8a}) we get
\begin{align}
\dot{p}_{0}&=\dot{p}_{2}=\dot{p}_{3}=0\,,\nonumber\\
e^{p_{0}/\kappa}f^{-1}\dot{p}_{1}&=f'p_{0}\dot{t}+e^{p_{0}/\kappa}p_{3}\dot{\phi}\,,\label{12}
\end{align}
while from (\ref{8b})
\begin{align}
f\dot{t}&=\frac{1}{\kappa}e^{p_{0}/\kappa}rf^{-1}\dot{p}_{1}+N\frac{\partial{\cal C}\left(p\right)}{\partial p_{0}}\,,\nonumber\\
\dot{\phi}&=\frac{e^{-p_{0}/\kappa}}{r}N\frac{\partial{\cal
C}\left(p\right)}{\partial p_{3}}\,,\label{13}\\
&\frac{\partial{\cal C}\left(p\right)}{\partial
p_{1}}=\frac{\partial{\cal C}\left(p\right)}{\partial
p_{2}}=0\,.\nonumber
\end{align}
Since $\partial{\cal C}\left(p\right)/{\partial p_{i}} \sim p_i$
this last equation tells that $p_1=p_2=0$.

Solving for $\dot\phi$ and $\dot t$ we obtain an equation for
components of momentum
\begin{equation}
p_{3}\frac{\partial{\cal C}\left(p\right)}{\partial
p_{3}}=-rf'f^{-1}p_{0}\frac{\partial{\cal C}\left(p\right)}{\partial
p_{0}}.\label{14}
\end{equation}
We expect that for physical bodies on the orbit in question we can
safely expand ${\cal C}$ in (\ref{10}) to the next-to-leading order,
to wit
$$
{\cal C}(p)= 1+ \frac{p^2_0}{2\kappa^2} -
\frac{\mathbf{p}^2}{2\kappa^2}\left(1+\frac{p_0}{\kappa}\right)+\ldots
$$
so that eq.\ (\ref{14}) takes the form
\begin{equation}
p_{3}^{2}=rf'f^{-1}p_{0}^{2}\left(1-\frac{1}{\kappa}p_{0}\left(1+\frac{1}{2}rf'f^{-1}\right)\right)\,.\label{15}
\end{equation}

We can now derive the first order in $1/\kappa$ corrections to the
angular velocity. Substituting Eq.~(\ref{15}) in the second of
Eqs.~(\ref{12}), one finds
\begin{equation}
 \frac{d\phi}{dt} =
\sqrt{\frac{ff'}{r}}\left(1-\frac{1}{2\kappa}p_{0}\left(1-\frac{1}{2}rf'f^{-1}\right)\right).
 \label{16}
\end{equation}

As it is well known from previous studies on relative locality~\cite{bob,k-bob,DSR-DeSitter}, the nontriviality (momentum dependence) in the (deformed-)relativistic transformations of spacetime coordinates imply that events distant from the origin of a given observer's frame are characterized by ``misleading inferences'' regarding their position in spacetime. The relative locality perspective is then that the physical observations corresponds to the ones made by observers local to the event of measurement~\cite{bob}. If one restricts to purely translated observers, one can ascribe the presence of these misleading inferences to the fact that the Poisson brackets between spacetime coordinates and momenta depend on momenta. Then the equations of motion are not sufficient for establishing the physical motion of particles, but one has to consider also the non trivial properties of translations~\cite{k-bob,DSR-DeSitter}. 
Even though it has been shown in previous studies such as Ref.~\cite{bob} that, if one considers the whole set of relativistic transformations (so to include also boosted observers), one cannot remove the effects of relative locality by a coordinate transformation (relative locality is not a coordinate artifact), it is still interesting to ask ourselves if it is possible to choose spacetime coordinates such that purely translated observers agree on the locality of distant events. This amounts to look for spacetime coordinates which have canonical Poisson brackets with momenta.

Of course, in the framework proposed in this manuscript, where both momentum space and spacetime curvature is present, the fact that one cannot rely in general on a full set of relativistic spacetime transformations, could require a generalization and/or revisal of these arguments. Nevertheless, postponing a more detailed investigation of the role of relative locality inferences in curved spacetime to a following work, we notice that from the action (\ref{5}), it is
straightforward to define the canonical variables
\begin{gather}
P_A = \left(E, P_r , P_\theta, L \right) = \left(fp_0 , f^{-1}p_1, r
p_2,
r p_3\right), \nonumber \\
X^A = \left( T, R, \Theta, \Phi\right) = \left( t, e^{p_0/\kappa} r,
e^{p_0/\kappa} \theta, e^{p_0/\kappa} \phi\right), \label{17}
\end{gather}
such that $\left\{ P_A ,X^B\right\}=\delta_A^B$. Notice that $E$ and
$L$ are respectively the particle's energy and angular momentum,
which are conserved even relaxing the circular orbits condition
($\dot{r}=0$).
 Substituting the canonical variables (\ref{17}) into Eq. (\ref{16}),
one finds the angular velocity
\begin{equation}
\frac{d\Phi}{dT} =
\sqrt{\frac{\delta\left(R\right)}{R^{2}}}\left(1+\frac{E}{4\kappa}\frac{2-3\delta\left(R\right)}{\left(1-2\delta\left(R\right)\right)^{3/2}}\right),
\label{AngVelocity}
\end{equation}
where $\delta(R)$ is the adimensional quantity $\delta(R) =
M\left(R\right)G / c^{2}R$, with $M(R)$ the mass
interior\footnote{To derive the relation between $\delta(r)$ and
$\delta(R)$ we have assumed spherical mass distribution, with
constant average density, so that $M(r)$ scales as $r^3$, i.e. $M(r)
= (M_{tot}/R^3_{tot})r^3 = e^{-3p_0/\kappa}(M_{tot}/R^3_{tot})R^3
=e^{-3p_0/\kappa}M(R)$.} to radius $R$.

Let us estimate the magnitude of the $\kappa$ correction term in
(\ref{AngVelocity}). Clearly, if the body in question is a planet in
the planetary system or a star in the binary system or the galaxy,
if we take $\kappa \sim M_{Pl}$ the correction $E/\kappa$ will be
enormously large. This is nothing but the `soccer ball problem' (see
\cite{AmelinoCamelia:2011uk} and references therein), and to get the
right size of the correction term we must invoke the way this
apparent paradox is solved. Namely, it is observed that the
interactions of a composite body (which, no doubts, the planet or
the star is) are governed by the deformation scale $\kappa_{eff}$
which is $N$ times bigger than the deformation scale for the
constituents $\kappa_{const}$, with $N$ being the number of
constituents. Therefore $\kappa$ in (\ref{AngVelocity}) is the
effective value of the deformation parameter, which equals, for
example $\kappa=N_{atoms}\, \kappa_{atom}$, with $\kappa_{atom}$
being the deformation parameter at the atomic scale. Then, since we
can safely assume in the present context that the moving body is
non-relativistic $E \sim N_{atoms}\,m_{atom}$. Therefore the ratio
$p_0/\kappa$ in (\ref{AngVelocity}) is of order of
$m_{atom}/\kappa_{atom}$, which, in the best case $\kappa_{atom}\sim
M_{Pl}$ is of order of $10^{-19}$, but might be few orders of
magnitude smaller than that.

We could ask if the terms multiplying the $\kappa$ correction in
(\ref{AngVelocity}) would act as amplifier for the smallness of the
effect. Taking $M_{tot}\sim10^{12}\text{ M}_{\odot}$, which is a
typical order of magnitude for large mass galaxies, and
$R\sim10^{3}\text{pc}$, we get $\delta (R)\sim10^{-7}$. In view of
this the corrections to the equation (\ref{AngVelocity}) resulting
from the non-trivial momentum space geometry, in the
non-relativistic regime are negligibly small.

In this paper we presented the theory of motion
of a relativistic particle with curved momentum
space traveling in a spacetime of an arbitrary geometry. 
There are many interesting physical situations where this theory 
can be used. The first is the cosmological setup, with spacetime being
of the form of de Sitter or a general Friedmann-Robertson-Walker,
the second the black hole geometry and/or Rindler space. In both
cases it is expected that the remnants of quantum gravity
effects might be relevant, and those could be described by
the framework presented in this paper.
Still, for each of these cases, one has to deal with the effects of 
relative locality, which are complementary to curvature 
of momentum space.
We will address these questions in the future publications.

\section*{Acknowledgements}

For JKG this work was supported in parts by the grant
2011/01/B/ST2/03354 and for JKG and GR by funds provided by the
National Science Center under the agreement DEC-
2011/02/A/ST2/00294.

\end{document}